\documentclass[aps,prl,twocolumn]{revtex4}
\usepackage{amsmath,amssymb}
\usepackage{graphicx}

\newcommand{\be}{\begin{equation}}
\newcommand{\ee}{\end{equation}}
\newcommand{\ba}{\begin{eqnarray}}
\newcommand{\ea}{\end{eqnarray}}
\newcommand{\ban}{\begin{eqnarray*}}
\newcommand{\ean}{\end{eqnarray*}}
\newcommand \nn {\nonumber}

\begin{document}

\title{$^4{\rm He}$ vs. $^4{\rm Li}$ and production of light nuclei in relativistic heavy-ion collisions}

\author{Sylwia Bazak$^1$\footnote{e-mail:sylwia.bazak@gmail.com} and Stanis\l aw Mr\' owczy\' nski$^{1,2}$\footnote{e-mail: MrowczynskiS@ncbj.gov.pl}}

\affiliation{$^1$Institute of Physics, Jan Kochanowski University, ul. \'Swi\c etokrzyska 15, 25-406 Kielce, Poland 
\\
$^2$National Centre for Nuclear Research, ul. Ho\. za 69, 00-681 Warsaw, Poland}

\date{June 18, 2018}

\begin{abstract}

We propose to measure the yields of $^4{\rm He}$ and $^4{\rm Li}$ in relativistic heavy-ion collisions to clarify a  mechanism of light nuclei production. Since the masses of $^4{\rm He}$ and $^4{\rm Li}$ are almost equal, the yield of $^4{\rm Li}$ predicted by the thermal model is 5 times bigger than that of $^4{\rm He}$ which reflects the different numbers of internal degrees of freedom of the two nuclides. Their internal structure is, however, very different: the alpha particle is well bound and compact while $^4{\rm Li}$ is weakly bound and loose. Within the coalescence model the ratio of yields of $^4{\rm Li}$ to $^4{\rm He}$ is shown to be significantly smaller than that in the thermal model and the ratio decreases fast from central to peripheral collisions of relativistic heavy-ion collisions because the coalescence rate strongly depends on the nucleon source radius. Since the nuclide $^4{\rm Li}$ is unstable and it decays into $^3{\rm He}$ and $p$ after roughly $30~{\rm fm}/c$, the yield of $^4{\rm Li}$ can be experimentally obtained through a measurement of the $^3{\rm He}\!-\!p$ correlation function. 

\end{abstract}


%Relativistic heavy-ion collisions, 25.75.−q
%Thermal and statistical models, 24.10.Pa  

\maketitle

One usually assumes that light nuclei are formed at the latest stage of relativistic heavy-ion collisions when a fireball disintegrates into hadrons which are flying away and are interacting only with their close neighbors in the phase-space. The final state interactions among nucleons are thus expected to be responsible for a production of light nuclei. This is the physical picture behind the coalescence model \cite{Butler:1963pp,Schwarzschild:1963zz} invented over half a century ago. We do not consider here nuclear fragments which are remnants of incoming nuclei formed out of spectator nucleons.  

The coalescence model is known to work well in a broad range of collision energies and  thus it is of no surprise that the model properly describes production of light nuclei and antinuclei in Pb-Pb collisions at $\sqrt{s_{\rm N\! N}}=2.76$ TeV \cite{Sun:2015ulc,Sun:2017ooe,Zhu:2015voa,Zhu:2017zlb,Wang:2017smh} which has been recently studied at LHC \cite{Adam:2015vda,Adam:2015yta,Acharya:2017bso}.

However, it has been recently observed that the yields of light nuclei and hypernuclei together with all other hadron species measured at LHC \cite{Adam:2015vda,Adam:2015yta,Acharya:2017bso} are also accurately described by the thermodynamical model \cite{Andronic:2010qu,Cleymans:2011pe} with a unique temperature of 156 MeV and vanishing baryon chemical potential relevant for midrapidity region of LHC. Simplicity of the thermal model makes its success very impressive but the result is truly surprising. It is hard to imagine that nuclei can exist in a hot and dense fireball. The temperature is much bigger than the nuclear binding energies and the inter-particle spacing is smaller than the typical size of light nuclei of interest. Therefore, the thermal model proponents speculate \cite{Andronic:2017pug} that the final state nuclei emerge from compact colorless droplets of quark-gluon matter already present in the fireball.

The thermal and coalescence models, which are physically quite different, were observed long ago to give rather similar yields of light nuclei \cite{Baltz:1993jh}, and recently the observation has been substantiated \cite{Zhu:2015voa,Mrowczynski:2016xqm} with a refined quantum-mechanical version of the coalescence model \cite{Sato:1981ez,Gyulassy:1982pe,Mrowczynski:1987,Lyuboshitz:1988,Mrowczynski:1992gc}. The question thus arises whether the final state formation of light nuclei can be quantitatively distinguished from the creation in a fireball. In other words, one asks whether the thermal approach to the production of light nuclei can be falsified. 

One of us suggested \cite{Mrowczynski:2016xqm} to compare the yield of $^4{\rm He}$, which was measured in relativistic-heavy ion collisions both at RHIC \cite{Agakishiev:2011ib} and LHC \cite{Adam:2015vda}, to the yield of exotic nuclide $^4{\rm Li}$ which was discovered in Brekeley in 1965 \cite{Cerny-1965}. The nuclide has spin 2 and it decays into $^3{\rm He}+p$ with the width of 6 MeV \cite{NNDC}, see also \cite{Tilley:1992zz}. The yield of $^4{\rm Li}$ can be experimentally obtained through a measurement of the $^3{\rm He}\!-\!p$ correlation function \cite{Pochodzalla:1987zz}. The alpha particle is well bound and compact while the nuclide $^4{\rm Li}$ is weakly bound and loose. Since the mass of $^4{\rm He}$ is smaller than that of $^4{\rm Li}$ by only 20 MeV, the yield of $^4{\rm Li}$ is according to the thermal model about five times bigger than that of $^4{\rm He}$ because of five spin states of $^4{\rm Li}$ and only one of $^4{\rm He}$. The aim of this note is to show that the coalescence model predicts a significantly smaller yield of $^4{\rm Li}$ due to its loose structure. 

The momentum distribution of a final state nucleus of $A$ nucleons is expressed in the coalescence model through the nucleon momentum distribution as
\be
\label{A-mom-dis}
\frac{dN_A}{d^3{\bf p}_A} = W \bigg(\frac{dN_N}{d^3{\bf p}} \bigg)^A,
\ee
where ${\bf p}_A = A{\bf p}$ and ${\bf p}$ is assumed to be much bigger than the characteristic momentum of a nucleon in the nucleus of interest. The coalescence formation rate $W$, which was first derived in \cite{Sato:1981ez} and later on repeatedly discussed \cite{Gyulassy:1982pe,Mrowczynski:1987,Lyuboshitz:1988,Mrowczynski:1992gc}, can be  approximated as
\ba \nn
\label{A-form-rate}
W &=& g_S g_I (2\pi)^{3(A-1)} V
\int d^3r_1 \, d^3r_2 \dots d^3r_A 
\\
&\times&
D ({\bf r}_1)\, D ({\bf r}_2) \dots D ({\bf r}_A)\, |\Psi({\bf r}_1,{\bf r}_2, \dots {\bf r}_A ) |^2 ,
\ea
where $g_S$ and $g_I$ are the spin and isospin factors to be discussed later on; the multiplier $(2\pi)^{3(A-1)}$ results from our choice of natural units where $\hbar =1$; $V$ is the normalization volume which disappears from the final formula; the source function $D ({\bf r})$ is the normalized to unity position distribution of a single nucleon at the kinetic freeze-out and $\Psi({\bf r}_1,{\bf r}_2, \dots {\bf r}_A )$ is the wave function of the nucleus of interest. The formula (\ref{A-mom-dis}) does not assume, as one might think, that the nucleons are emitted simultaneously. The vectors ${\bf r}_i$ with $i=1,2, \dots A$ denote the nucleon positions at the moment when the last nucleon is emitted from the fireball. For this reason, the function $D ({\bf r}_i)$ actually gives the space-time distribution and it is usually assumed to be Gaussian. We choose the isotropic form
\be
\label{Gauss-source}
D ({\bf r}_i) = (2 \pi R_s^2)^{-3/2} \, e^{-\frac{{\bf r}_i^2}{2R^2_s}},
\ee
where $R_s$ is the root mean square (RMS) radius of the nucleon source, because the coalescence rate does not allow one to disentangle the source radii in different directions as they enter the rate in the combination which is independent of a momentum of the nucleus under consideration. 

To formulate a relativistically covariant coalescence model one usually uses the Lorentz invariant nucleon momentum distributions in the relation analogous to (\ref{A-mom-dis}) and modifies the coalescence rate formula (\ref{A-form-rate}), see e.g. \cite{Sato:1981ez,Mrowczynski:1987}. Since we are interested in the ratio of the coalescence rates of $^4{\rm Li}$ and $^4{\rm He}$, our final result is insensitive to these heuristic modifications which are anyway not well established, as the relativistic theory of strongly interacting bound states is not fully developed. 

The modulus squared of the wave function of $^4{\rm He}$ is chosen as
\be
\label{alpha-wave-fun}
|\Psi_{\rm He}({\bf r}_1,{\bf r}_2, {\bf r}_3, {\bf r}_4) |^2 = C_\alpha 
e^{- \alpha ({\bf r}_{12}^2 + {\bf r}_{13}^2 + {\bf r}_{14}^2 + {\bf r}_{23}^2 + {\bf r}_{24}^2 + {\bf r}_{34}^2)},
\ee
where $C_\alpha$ is the normalization constant,  ${\bf r}_{ij} \equiv {\bf r}_i - {\bf r}_j$ and $\alpha$ is the parameter to be related to the RMS radius of $^4{\rm He}$ which is denoted as $R_\alpha$. We further use the Jacobi variables defined as
\ba
\nn
{\bf R} &\equiv& \frac{1}{4}({\bf r}_1 + {\bf r}_2 + {\bf r}_3 + {\bf r}_4) ,
\\ \nn
{\bf x} &\equiv& {\bf r}_2 - {\bf r}_1 ,
\\ \nn
{\bf y} &\equiv& {\bf r}_3 - \frac{1}{2}({\bf r}_1 + {\bf r}_2 ) ,
\\ 
\label{Jacobi-z}
{\bf z} &\equiv& {\bf r}_4 - \frac{1}{3}({\bf r}_1 + {\bf r}_2 + {\bf r}_3) ,
\ea
which have the nice property that the sum of squares of particles' positions and the sum of squares of differences of the positions are expressed with no mixed terms of the Jacobi variables that is
\ba
\nn
{\bf r}_1^2 + {\bf r}_2^2 + {\bf r}_3^2 + {\bf r}_4^2 
= 4 {\bf R}^2 + \frac{1}{2} {\bf x}^2 + \frac{2}{3} {\bf y}^2 + \frac{3}{4} {\bf z}^2 ,
\\ \nn
{\bf r}_{12}^2 + {\bf r}_{13}^2 + {\bf r}_{14}^2 + {\bf r}_{23}^2 + {\bf r}_{24}^2 + {\bf r}_{34}^2 = 2 {\bf x}^2 + \frac{8}{3} {\bf y}^2 + 3 {\bf z}^2 .
\ea
Then, one easily finds that 
\be
C_\alpha = \frac{2^6}{V} \Big(\frac{\alpha}{\pi}\Big)^{9/2}, 
~~~~~~~~~~ 
\alpha = \frac{3^2}{2^5 R_\alpha^2}  .
\ee

Substituting the formulas (\ref{Gauss-source}) and (\ref{alpha-wave-fun}) into Eq.~(\ref{A-form-rate}), one finds the coalescence rate of  $^4{\rm He}$ as
\be
\label{alpha-rate}
W_{\rm He} =  \frac{\pi^{9/2}}{2^{9/2}}
\frac{1}{\big(R_s^2 +\frac{4}{9} R_\alpha^2\big)^{9/2}} ,
\ee 
where the spin and isospin factors have been included. Since $^4{\rm He}$ is the state of zero spin and zero isospin, the factors are 
\be
g_S = g_I = \frac{1}{2^3}, 
\ee
because there are $2^4$ spin and $2^4$ isospin states of four nucleons and there are two zero spin and two zero isospin states. The coalescence rate of $^4{\rm He}$ was computed long ago in \cite{Sato:1981ez}.

The stable isotope $^6{\rm Li}$ is a mixture of two cluster configurations $ ^4{\rm He}\!-\! ^2{\rm H}$ and $ ^3{\rm He}\!-\! ^3{\rm H}$ \cite{Bergstrom:1979gpv}. Since $^4{\rm Li}$ decays into $^3{\rm He}+p$, we assume that it has the cluster structure $ ^3{\rm He}\!-\!p$ and following \cite{Bergstrom:1979gpv} we parametrize the modulus squared of the wave function of $^4{\rm Li}$ as
\ba
\nn
|\Psi_{\rm Li}({\bf r}_1,{\bf r}_2, {\bf r}_3, {\bf r}_4) |^2 = 
C_{\rm Li} \, e^{-\beta ({\bf r}_{12}^2 + {\bf r}_{13}^2 + {\bf r}_{23}^2 )} 
\\
\label{Li-wave-fun}
\times ~
{\bf z}^4 e^{-\gamma {\bf z}^2} \, |Y_{lm}(\Omega_{\bf z})|^2,
\ea
where the nucleons number 1, 2 and 3 form the $^3{\rm He}$ cluster while the nucleon number 4 is the proton; ${\bf z}$ is the Jacobi  variable (\ref{Jacobi-z}); $Y_{lm}(\Omega_{\bf z})$ is the spherical harmonics related to the rotation of the vector ${\bf z}$ with quantum numbers $l,m$. The summation over $m$ is included in the spin factor $g_S$. Using the Jacobi variables, one analytically computes the normalization constant $C_{\rm Li}$ and expresses the parameter $\beta$ through the RMS radius $R_c$ of the cluster $^3{\rm He}$ as
\be
C_{\rm Li} = \frac{2^4 3^{1/2} \beta^3 \gamma^{7/2}}{5 \pi^{7/2} V}, ~~~~~~~~~ 
\beta = \frac{1}{3 R_c^2}  .
\ee
The parameter $\gamma$  is expressed through the RMS radius $R_{\rm Li}$ of $^4{\rm Li}$ and the cluster radius $R_c$ in the following way
\be
\gamma = \frac{21}{2^3(4R_{\rm Li}^2 - 3R_c^2)} .
\ee

Let us now discuss the spin and isopsin factors which enter the coalescence rate of  $^4{\rm Li}$. The nuclide has the isospin $I =1,~I_z = 1$ and thus the isospin factor is
\be
\label{gI-Li}
g_I = \frac{3}{2^4},
\ee
because there are three isospin states $I =1,~I_z = 1$ of four nucleons. The spin of $^4{\rm Li}$ is 2 but we do not know what is the orbital contribution. The spin 2 of  $^3{\rm He}$ and $p$ can be arranged with the orbital angular momentum $l=1$ and $l=2$. We assume here that the cluster $^3{\rm He}$ is of spin $1/2$ as the free nuclide  $^3{\rm He}$. (If the spin $3/2$ of $^3{\rm He}$ were allowed, the orbital number $l=0$ would be also possible.) When $l=2$, the total spin of $^3{\rm He}$ and $p$ has to be zero and thus
\be
\label{gS-l2}
g_S = \frac{1}{2^3} .
\ee
If $l=1$, the total spin of $^3{\rm He}$ and $p$ has to be one and there are $3^2$ such spin states of four nucleons. Consequently, there are $3^2$  angular momentum states with 5 states corresponding to spin 2 of  $^4{\rm Li}$ and thus
\be
\label{gS-l1}
g_S = \frac{3^2}{2^4} \frac{5}{3^2} = \frac{5}{2^4}.
\ee

Substituting the formulas (\ref{Gauss-source}) and (\ref{Li-wave-fun}) into Eq.~(\ref{A-form-rate}), one finds the coalescence rate of  $^4{\rm Li}$ as
\be
\label{Li-rate}
W_{\rm Li} 
= \frac{3 \pi^{9/2}}{2^{11/2}} {\frac{5}{2} \choose 1}
 \frac{R_s^4}{\big(R_s^2 +\frac{1}{2} R_c^2\big)^3 
\big(R_s^2 +\frac{4}{7} R_{\rm Li}^2 - \frac{3}{7} R_c^2\big)^{7/2}} ,
\ee 
where the upper case is for $l=1$ and the lower one for $l=2$. Since the source function (\ref{Gauss-source}) is spherically symmetric, the coalescence rate (\ref{Li-rate}) depends on the orbital numbers $l$ only through the spin factor $g_S$. 

\begin{figure}[t]
\centering
\includegraphics[scale=0.275]{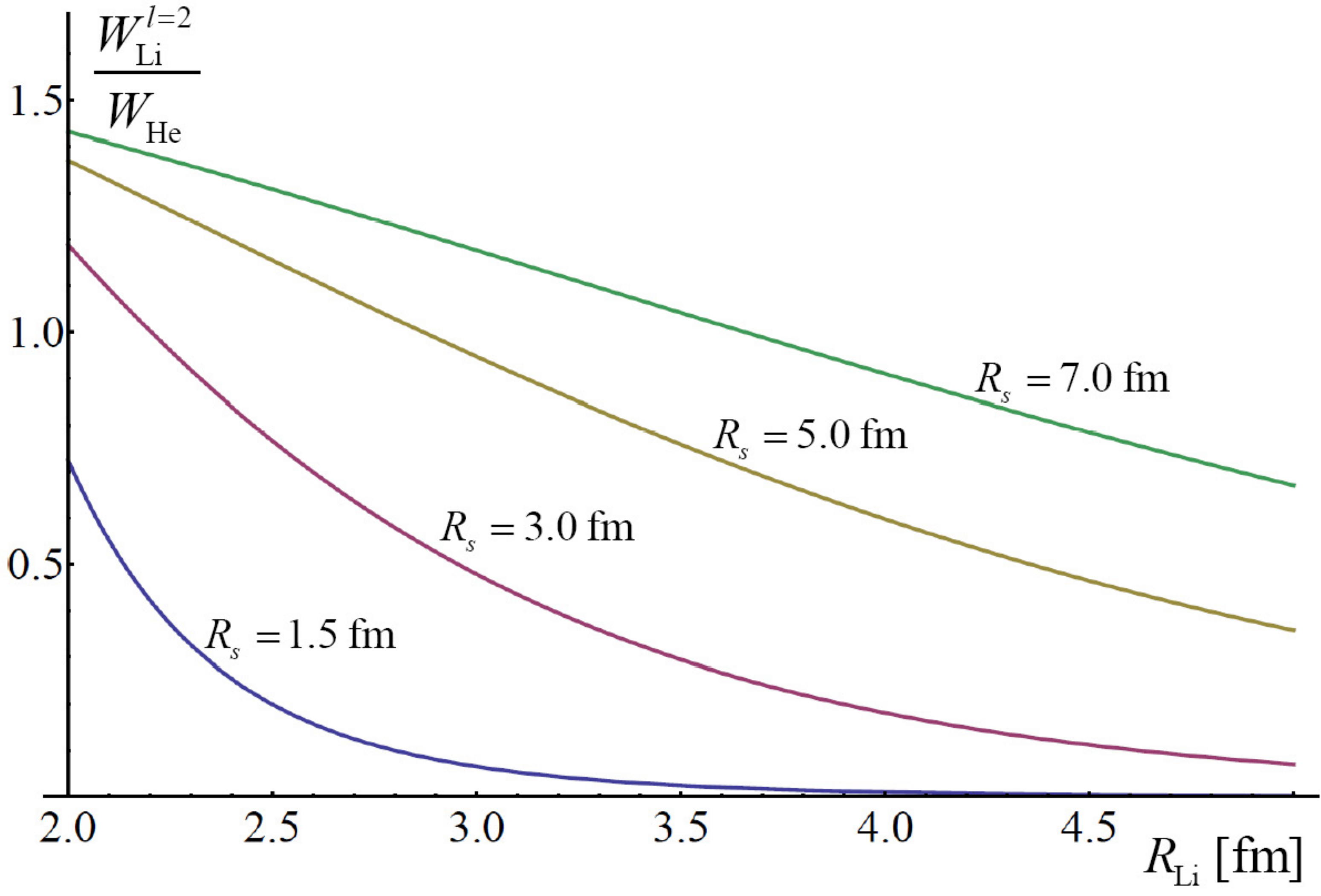}
\vspace{-7mm}
\caption{The ratio of formation rates of $^4{\rm Li}$ in $l=2$ state and $^4{\rm He}$ as a function of $R_{\rm Li}$ for four values of $R_s = 1.5,~3.0,~5.0$ and 7.0 fm.}
\label{Fig1}
\end{figure}

We note that even when $R_\alpha = R_{\rm Li}$ and the spin-isospin factors are ignored, the coalescence rates of $^4{\rm He}$ and   $^4{\rm Li}$ still differ from each other because the internal structure of $^4{\rm He}$ differs from that of  $^4{\rm Li}$. The rates become equal when $R_s \gg R_\alpha $ and $R_s \gg R_{\rm Li}$ as then the structure of nuclei does not matter any more. One checks that our  formulas indeed confirm the expectation. 

The ratio of yields of $^4{\rm Li}$ and $^4{\rm He}$ is given by the ratio of the formation rates $W_{\rm Li}$ and $W_{\rm He}$. The latter ratio depends on four parameters: $R_s$, $R_\alpha$, $R_{\rm Li}$ and $R_c$. The fireball radius at the kinetic freeze-out $R_s$ is determined by the femtoscopic $\pi\!\!-\!\!\pi$ correlations. Specifically, the experimentally measured radii $R_{\rm out},\, R_{\rm side},\, R_{\rm long}$ can be used to get the kinetic freeze-out radius as $R_s = (R_{\rm out} R_{\rm side} R_{\rm long})^{1/3}$. Then, the source radius $R_s$ varies from peripheral to central Pb-Pb collisions at LHC between, say, 3 and 7 fm \cite{Adam:2015vna}. The RMS radius of $^4{\rm He}$ is $R_\alpha = 1.68$ fm \cite{Angeli:2013epw} and the RMS radius of the cluster $^3{\rm He}$ is identified with the radius of a free nucleus $^3{\rm He}$ and thus $R_c = 1.97$ fm \cite{Angeli:2013epw}. The radius $R_{\rm Li}$ is unknown but obviously it must be bigger than $R_c$. Taking into account a finite size of a proton it is fair to expect that $R_{\rm Li}$ is at least 2.5--3.0 fm. The ratio of the formation rates $W_{\rm Li}^{l=2}$ and $W_{\rm He}$ is shown in Fig.~\ref{Fig1} as a function of $R_{\rm Li}$ for four values of $R_s = 1.5,~3.0,~5.0$ and 7.0 fm. The ratio of $W_{\rm Li}^{l=1}$ to $W_{\rm He}$ is bigger by the factor $5/2$. 

As already mentioned, the ratio of yields of $^4{\rm Li}$ and $^4{\rm He}$ equals 5 according to the thermal model. One sees in Fig.~\ref{Fig1} that the ratio is significantly smaller in the coalescence model. For $R_{\rm Li}= 3$ fm and the most central collisions of the heaviest nuclei, which corresponds to  $R_s \approx 7$ fm, the ratio $W_{\rm Li}^{l=2}/W_{\rm He}$ equals 1.2 but it drops to 0.7 for the centrality of 40-60\% where $R_s \approx 4$ fm. When $l=1$, the numbers are bigger by the factor $5/2$. The strong dependence of the yields of $^4{\rm Li}$ to $^4{\rm He}$ on the collision centrality is a characteristic feature of the coalescence mechanism because the coalescence rate decreases fast when the nucleon source radius goes to zero. Therefore, it should be possible to quantitatively distinguish the coalescence mechanism of light nuclei production from the creation in a fireball. 

The yield of $^4{\rm Li}$ can be experimentally obtained through a measurement of the $^3{\rm He}\!-\!p$ correlation function at small relative momenta. Such a measurement was successfully preformed in $^{40}{\rm Ar}-$induced reactions on $^{197}{\rm Au}$ at the collision energy per nucleon of 60 MeV \cite{Pochodzalla:1987zz}. The $^3{\rm He}\!-\!p$ correlation function is presented in Fig.~6 of Ref. \cite{Pochodzalla:1987zz} and the peak of $^4{\rm Li}$ is clearly seen. The proposed measurement in relativistic heavy-ion collisions at LHC is challenging but possible \cite{private} and we just work on the theoretically expected $^3{\rm He}\!-\!p$ correlation function. The problem, however, is not simple: these are not $^3{\rm He}$ and $p$ but four nucleons, which are emitted from a source, and their wave function should be projected on the correlated state of $^3{\rm He}$ and $p$. We deal here with a rather complex coupled channel problem where several angular momenta and isopin states must be considered. 

\vspace{5mm}

The authors gratefully acknowledge discussions and correspondence with P. Braun-Munzinger, W. Broniowski,  J. Schukraft and S. Wycech.

\end{document}